\documentclass{article}
\usepackage{newtxtext}
\usepackage{newtxmath}
\usepackage{booktabs}
\usepackage{tabularx}
\usepackage{ragged2e}
\usepackage{spconf,amsmath,graphicx,hyperref}
\usepackage{algorithm}
\usepackage{algorithmic}
\usepackage{multirow}
\usepackage{booktabs}
\usepackage{amsmath}
\usepackage{xcolor}
\usepackage{amsmath}
\usepackage{amsfonts}

\title{RRPO: ROBUST REWARD POLICY OPTIMIZATION \\FOR LLM-BASED EMOTIONAL TTS}

\name{
\begin{tabular}{@{}c@{}}
    Cong Wang$^{\dagger}$
    \quad Changfeng Gao$^{1}$ 
    \quad Yang Xiang$^{1}$
    \quad Zhihao Du$^{1}$    
    \quad Keyu An$^{1}$    
    \\
    \quad \textit{Han Zhao}$^{1}$  
    \quad \textit{Qian Chen}$^{1}$    
    \quad \textit{Xiangang Li}$^{1}$          
    \quad \textit{Yingming Gao}
    \quad \textit{Ya Li}$^{*}$
    \thanks{$^*$ Corresponding author. Email: yli01@bupt.edu.cn}
    \thanks{$^\dag$ Work performed during the internship at Tongyi Lab.}
    \end{tabular}
    }
    
\address{
    $^{1}$Speech Team, Tongyi Lab, Alibaba Group\\
    congwang@bupt.edu.cn
    }
\begin{document}
\ninept
\maketitle

\begin{abstract}
Differentiable reinforcement learning (RL) frameworks like DiffRO offer a powerful approach for controllable text-to-speech (TTS), but are vulnerable to reward hacking, particularly for nuanced tasks like emotion control. 
The policy model can exploit a vanilla Reward Model (RM) by generating acoustic artifacts to achieve spurious rewards, but at the cost of degrading perceptual quality. 
To address this, we propose Robust Reward Policy Optimization (RRPO), a novel framework that employs a hybrid regularization scheme. 
This scheme develops a robust RM whose reward signal is more reliably aligned with human perception, compelling the policy to abandon detrimental shortcuts and instead learn the complex features of genuine emotions. 
Our ablation study confirms the enhanced robustness of our RM, as evidenced by its strong cross-lingual generalization. 
The subjective evaluation demonstrates that this robust RM effectively mitigates reward hacking, leading to significant improvements in both emotional expressiveness and naturalness over all baselines. 
Demo page: https://lrwinr.github.io/RRPO-CosyVoice.
\end{abstract}

\section{Introduction}
\label{sec:introduction}
The field of text-to-speech (TTS) has been fundamentally reshaped by the advent of Large Language Models (LLMs), establishing neural codec language models (LMs) as the leading approach for high-quality synthesis~\cite{cosyvoice2, seedtts}. 
By leveraging the in-context learning capabilities of LLMs, these systems have achieved unprecedented quality and naturalness, enabling versatile applications like zero-shot voice cloning and instruction-based synthesis. 
A vital area in this domain is controllable emotional TTS, which must generate speech that is both emotionally accurate and highly expressive to be effective in human-computer interaction. 
However, the focus of current methods is often limited to categorical accuracy, while achieving the rich \textit{expressiveness} of human speech remains a significant challenge. 

Prior work has primarily approached this challenge from two directions. 
The first involves Supervised Fine-Tuning (SFT) on emotional datasets, using either categorical labels~\cite{emodiff,zetspeech,emomix} or natural language prompts~\cite{voxinstruct,emovoice,instructtts}. 
While effective for basic emotional control, SFT-based methods are fundamentally limited by the diversity of their training data. Consequently, they struggle to generate expressiveness beyond what is explicitly present in the corpus. 
The second, more recent direction uses Reinforcement Learning (RL) to achieve preference alignment in LLM-based TTS systems~\cite{preference, emodpo}. 
A significant advancement is Differentiable Reward Optimization (DiffRO)~\cite{DiffRO}, which formulates policy optimization as a fully differentiable process. 
By enabling gradients to be back-propagated from a multi-task Reward Model (RM) to the policy model (i.e., the LM), DiffRO avoids the high variance of traditional Policy Gradient (PG) methods~\cite{PPO, DPO, GRPO}. 

However, the direct, deterministic optimization of DiffRO, while being its primary strength, also introduces a significant vulnerability. 
The framework's effectiveness fundamentally depends on the quality of its RM. A vanilla RM can be easily exploited by the policy model, a phenomenon known as \textit{reward hacking}~\cite{rewardhacking}. 
We observe a critical issue with this approach: the policy model learns to take a detrimental ``shortcut" by generating non-semantic \textit{acoustic artifacts}, such as unnatural mouth clicks or harsh plosives, which successfully fool the RM into assigning a high reward but come at the cost of overall perceptual quality. 
This leads to a paradoxical outcome, where the policy model receives a higher reward from the RM, even as the human-perceived quality of its output, particularly naturalness and pronunciation, severely degrades. 
This vulnerability exposes a fundamental gap: without a robust reward signal, powerful optimization frameworks like DiffRO can inadvertently guide the policy away from the true goal of generating genuinely expressive speech. 

To address this challenge, we propose Robust Reward Policy Optimization (RRPO), a framework designed to mitigate \textit{reward hacking} and enhance the genuine \textit{emotional expressiveness} of TTS systems. 
We hypothesize that the solution to this problem depends on first building a robust RM that is reliably aligned with human perception. We achieve this through a novel hybrid regularization scheme that fine-tunes a pre-trained RM, correcting its vulnerabilities at three levels: overconfidence, brittle decision boundaries, and perturbation sensitivity. 
A robust RM is not easily fooled by simple acoustic artifacts, which in turn compels the policy model to abandon shortcuts and instead learn the more complex acoustic features that align with genuine human perception. 
Our main contributions are:
\begin{figure*}[!htb]
\begin{minipage}[b]{1.0\linewidth}
  \centering
  \centerline{\includegraphics[width=0.93\linewidth]{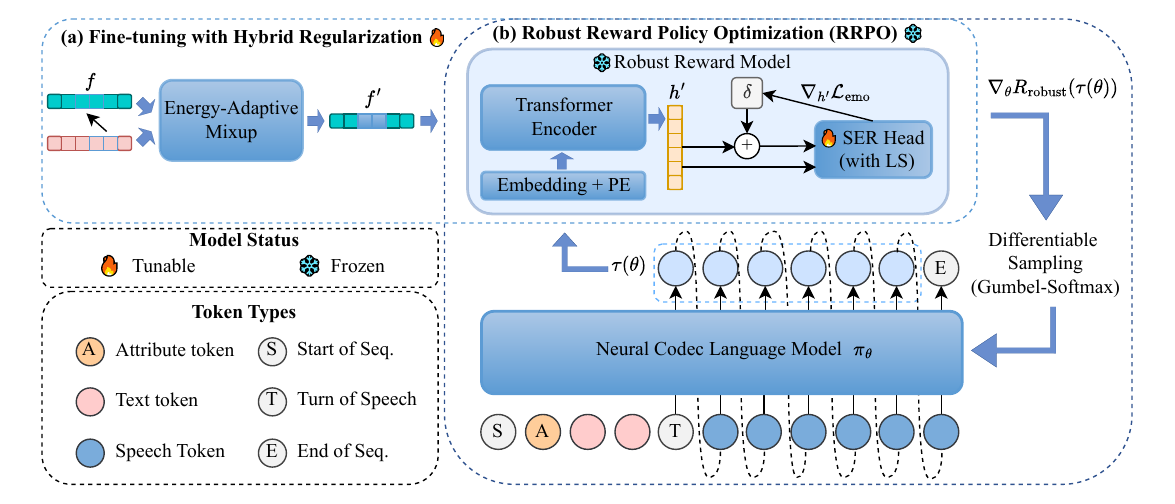}}
  \vspace{-0.5cm}
\end{minipage}
\caption{The framework of our proposed Robust Reward Policy Optimization (RRPO).}
\label{fig:overview}
\vspace{-0.2cm}
\end{figure*}
\begin{itemize}
    \item We identify and analyze a critical vulnerability in the DiffRO framework for emotional TTS: a reward hacking mechanism where the policy learns to generate acoustic artifacts to exploit the RM for spurious rewards. 
    \item We propose a novel hybrid regularization scheme that significantly enhances the robustness of the RM, as validated by an objective, cross-lingual SER task. 
    \item We demonstrate through subjective evaluation that RRPO effectively mitigates reward hacking, leading to substantial improvements in both emotional expressiveness and naturalness over strong baselines.
\end{itemize}

\section{Methodology}
\label{sec:methodology}

Fig.~\ref{fig:overview} illustrates the framework of our proposed method, RRPO, which enhances DiffRO~\cite{DiffRO} by employing a hybrid regularization fine-tuning scheme for the RM. 

\subsection{Motivation: The Need for a Robust Reward Model}
\label{ssec:motivation}

As introduced in Sec.~\ref{sec:introduction}, DiffRO~\cite{DiffRO} provides a powerful, low-variance alternative to traditional PG methods by establishing a fully differentiable optimization progress. 
This requires two key components: first, treating the generated trajectory $\tau$ as a differentiable function of the policy parameters $\theta$, $\tau(\theta)$, enabled by Gumbel-Softmax reparameterization~\cite{gumbelsoftmax}; and second, employing a differentiable RM. Together, these components enable the direct, analytical computation of the policy gradient via the chain rule: 
\begin{equation}
    \nabla_\theta J(\theta) = \nabla_\theta R(\tau(\theta)) = \frac{\partial R}{\partial \tau} \cdot \frac{\partial \tau}{\partial \theta}.
\end{equation} 
where, the reward $R(\tau(\theta))$ is defined as the negative of a differentiable loss function, such as one for Speech Emotion Recognition (SER). 
This analytical gradient provides a key advantage over methods like DPO and GRPO~\cite{DPO, GRPO}, which only estimate the gradient's direction. By providing both precise direction and magnitude, DiffRO enables highly efficient, token-level policy optimization. 

However, this high-efficiency optimization comes with a critical vulnerability. 
Any flaw or bias in the RM is amplified by the powerful analytical gradient, producing an incorrect gradient that can rapidly guide the policy toward a degenerate, exploitative solution. 
This susceptibility to reward hacking makes the RM's robustness a critical prerequisite for the framework to be effective and safe. 
Motivated by this necessity, the following section details our hybrid regularization scheme for building such a model.

\subsection{Fine-Tuning with Hybrid Regularization}
\label{ssec:fine-tuning}

However, a key challenge in RL is {\it reward hacking}, where the policy model exploits the vanilla RM to maximize spurious rewards at the expense of true alignment. 
To mitigate this vulnerability, we perform a robust fine-tuning process on the pre-trained RM using a hybrid regularization scheme. This process leverages a small, high-quality, human-annotated dataset to correct potential biases introduced during pre-training. And we integrate three complementary regularization techniques to prevent the RM from overfitting to this significantly smaller dataset and developing new biases. 

\vspace{-0.2cm}
\subsubsection{Label Smoothing: Correcting Overconfidence}
\label{sssec:labelsmoothing}

A primary bias the RM can exhibit is overconfidence in its predictions. This issue is exacerbated by the use of discrete categorical labels in the SER task, which fail to capture the continuous and often ambiguous nature of human emotion. 
To correct this overconfidence, we employ Label Smoothing (LS)\cite{labelsmoothing}, which replacing the hard, one-hot label $\mathbf{y}$ with a soft probability distribution $\mathbf{y}'$:

\begin{equation}
    y'_{k} = (1 - \epsilon) \cdot y_{k}~ + ~\frac{\epsilon}{K}.
\label{eq:labelsmoothing}
\end{equation} 
where $K$ is the number of classes and $\epsilon$ is a hyperparameter. By penalizing overconfidence, LS improves the RM's robustness. 

\vspace{-0.2cm}
\subsubsection{Energy-Adaptive Mixup: Correcting Brittle Decision Boundaries}
\label{sssec:eam}

Another cause of reward hacking is the RM's brittle decision boundaries, allowing the policy model to make small, perceptually meaningless modifications to the input that result in spurious high rewards. 
To correct this, we employ Energy-Adaptive Mixup (EAM), a data augmentation technique specifically designed for speech signals~\cite{labelsmoothing, Wavlm, LQF}. 
Unlike conventional Mixup~\cite{mixup}, EAM calculates a mixing coefficient $\lambda$ based on the relative energy and duration of the mixed speech segments. 
The final loss is an adaptively weighted interpolation of the LS losses ($\mathcal{L}_{\text{LS}}$) computed between the RM's prediction and each of the two original labels. 
This encourages the RM to learn a smooth transition between data points, thereby correcting sharp, brittle decision boundaries and making it significantly harder for the policy model to exploit such vulnerabilities. 
Specifically, we apply EAM to each batch of low-level acoustic features $\mathbf{F}$. For each sample $\mathbf{f}_i$ and its paired sample $\mathbf{f}_j$, EAM generates a mixed feature vector $\mathbf{f}'_i$ and a corresponding energy-adaptive mixing coefficient $\lambda_i$, as detailed in Algorithm~\ref{alg:eam}. These mixed features is then passed through a transformer encoder to produce a high-level embeddings $\mathbf{h}'$, from which the RM predicts the final output $\hat{\mathbf{y}}$. 
The loss is computed as: 
\vspace{-0.2cm}
\begin{equation}
    \mathcal{L}_{\text{emo}} = \frac{1}{B} \sum_{i=1}^{B} \left[ 
    (1 - \lambda_i)  \mathcal{L}_{\text{LS}}(\hat{y_i},~{y_i}') 
    + 
    \lambda_i \mathcal{L}_{\text{LS}}(\hat{y_i},~{y_j}') 
    \right],
\label{eq:mixed_loss}
\end{equation} 

\vspace{-0.5cm}
\begin{algorithm}[t!h]
\caption{Energy-Adaptive Mixup (EAM)}
\label{alg:eam}
\begin{algorithmic}[1]
\REQUIRE 
    \begin{minipage}[t]{0.9\linewidth} 
    A batch of features $\mathcal{F}_B = \{\mathbf{f}_i \in \mathbb{R}^{L_i \times D}\}_{i=1}^B$. \\
    Corresponding one-hot labels $\mathcal{Y}_B = \{y_i\}_{i=1}^B$. \\
    SNR(dB) mix ratio range $[r_{\min}, r_{\max}]$.
    \end{minipage}
\ENSURE
    \begin{minipage}[t]{0.85\linewidth} 
    The batch of mixed features $\mathcal{F}'_B = \{\mathbf{f}'_i\}_{i=1}^B$. \\
    The batch of paired labels $\mathcal{Y}_j = \{y_j\}_{i=1}^B$. \\
    The batch of mixing coefficients $\Lambda = \{\lambda_i\}_{i=1}^B$.
    \end{minipage}
    
\STATE Generate a random permutation $\pi$ of $\{1, \dots, B\}$
\STATE Initialize collections $\mathcal{F}'_B, \mathcal{Y}_j, \Lambda$ of size $B$

\FOR{$i = 1 \ \text{to}\  B$}
    \STATE $j \leftarrow \pi_i$
    \IF{$i == j$} 
        \STATE $j \leftarrow (j \pmod B) + 1$
    \ENDIF
    
    \STATE Let $l_i, l_j$ be the lengths of $\mathbf{f}_i, \mathbf{f}_j$
    \STATE Sample mix length: $l_{\text{mix}} \sim \mathcal{U}\left(1, \frac{l_i}{2}\right)$, then cast to integer
    
    \IF{$l_{\text{mix}} > l_j$}
        \STATE $l_{\text{mix}} \leftarrow l_j$
    \ENDIF
    
    \STATE Sample start indices: $b_i \sim \mathcal{U}\{0, \dots, l_i - l_{\text{mix}}\}$, $b_j \sim \mathcal{U}\{0, \dots, l_j - l_{\text{mix}}\}$
    
    \STATE Extract segments: 
           $\mathbf{s}_i \leftarrow \mathbf{f}_i[b_i : b_i + l_{\text{mix}}]$, 
           $\mathbf{s}_j \leftarrow \mathbf{f}_j[b_j : b_j + l_{\text{mix}}]$
           
    \STATE Calculate mean segment energies: 
           $E_i \leftarrow \text{mean}(\mathbf{s}_i^2)$, 
           $E_j \leftarrow \text{mean}(\mathbf{s}_j^2)$
           
    \STATE Sample a mix ratio: $r \sim \mathcal{U}[r_{\min}, r_{\max}]$
    \STATE Calculate target energy: $E_j' \leftarrow \frac{E_i}{10^{r / 10}}$
    \STATE Scale segment $\mathbf{s}_j$: 
           $\mathbf{s}_j' \leftarrow \sqrt{E_j' / E_j} \times \mathbf{s}_j$
           
    \STATE Create mixed feature by overlap-add: $\mathbf{f}'_i \leftarrow \mathbf{f}_i$
    \STATE $\mathbf{f}'_i[b_i : b_i + l_{\text{mix}}] \leftarrow \mathbf{f}'_i[b_i : b_i + l_{\text{mix}}] + \mathbf{s}_j'$
    
    \STATE Calculate the energy-adaptive mixing coefficient: $\lambda_i \leftarrow \frac{l_{\text{mix}}}{l_i} \times \frac{E_j'}{E_i + E_j'}$.
           
    \STATE Store results for the $i$-th sample: $\mathcal{F}'_B[i] \leftarrow \mathbf{f}'_i$, $\mathcal{Y}_j[i] \leftarrow \mathcal{Y}_B[j]$, $\Lambda[i] \leftarrow \lambda_i$
\ENDFOR

\RETURN $\mathcal{F}'_B$, $\mathcal{Y}_j$, $\Lambda$
\end{algorithmic}
\end{algorithm}

\vspace{-0.5cm}
\subsubsection{Adversarial Training: Correcting Perturbation Sensitivity}
\label{sssec:fgm}

A more subtle vulnerability of the RM is its sensitivity to minor perturbations. The policy model can learn to exploit this by introducing subtle distortions that fool the RM into assigning a higher reward. However, these perturbations degrade the perceptual quality of the speech. 
To correct this, we incorporate Adversarial Training (Adv). Specifically, we use a variant of the Fast Gradient Method~\cite{FGM}. The perturbation is applied directly to the high-level embeddings $\mathbf{h}'$. This targets the model's understanding of high-level emotional features rather than low-level acoustic features. It computes the worst-case perturbation $\mathbf{\delta}$ by ascending the normalized gradient of the loss: 
\vspace{-0.1cm}
\begin{equation}
    \mathbf{\delta} = \epsilon_{\text{adv}} \cdot \frac{\nabla_{\mathbf{h}'} \mathcal{L}_{\text{emo}}}{\|\nabla_{\mathbf{h}'} \mathcal{L}_{\text{emo}}\|_2},
\label{eq:fgm_perturbation}
\end{equation} 
where $\epsilon_{\text{adv}}$ is the perturbation magnitude. Adversarial samples are created by adding this to embeddings: $\mathbf{h}'_{\text{adv}} = \mathbf{h}' + \mathbf{\delta}$. These perturbed embeddings are then passed to compute adversarial loss, $\mathcal{L}_{\text{adv}}$, using the same formulation as Eq.~\ref{eq:mixed_loss}. 

\vspace{-0.2cm}
\subsubsection{The Final Corrective Objective}
\label{sssec:finalobjective}

This hybrid regularization scheme corrects the RM at three levels: label confidence, decision boundary, and perturbation sensitivity. The final SER loss is defined as: 
\begin{equation}
    \mathcal{L}_{\text{ser}} = \mathcal{L}_{\text{emo}} + \alpha \cdot \mathcal{L}_{\text{adv}},
\label{eq:final_loss}
\end{equation} 
where $\alpha$ is a balancing hyperparameter. 
By minimizing this loss, we develop a corrected RM that is significantly more robust against exploitation.
As a result, this enhanced robustness ensures the RM serves as a more reliable guide for the policy optimization phase, thereby mitigating the risk of reward hacking. 

\vspace{-0.2cm}
\subsection{Robust Reward Policy Optimization}
\label{ssec:rrpo}

Finally, the policy objective $J(\theta)$ is guided by our robust RM:
\begin{equation}
    \nabla_\theta J(\theta) = \nabla_\theta R_{\text{robust}}(\tau(\theta)),
\label{eq_rrpo_gradient}
\end{equation} 
where the robust reward $R_{\text{robust}}$ is the negative of hybrid-regularized SER loss, $-\mathcal{L}_{\text{ser}}$, from Eq.~\ref{eq:final_loss}. 
By using gradients from this robust reward, RRPO guides the policy with updates that are reliably aligned with true perceptual quality. This mitigates reward hacking and leads to a more controllable and expressive emotional TTS system. 

\section{Experiments}
\label{sec:experiments}

\subsection{Experimental Setup}
\label{ssec:setup}

\subsubsection{Baselines, Dataset and Implementation Details} 
\label{ssec:baselinesanddataset}

We compare our RRPO against three baselines. The first is the Cosyvoice2~\cite{cosyvoice2} model, which serves as our baseline system. The second is the SFT baseline, where the LM is fine-tuned on the target speaker's emotional dataset. The third is the DiffRO baseline~\cite{DiffRO}. 

For our experiments, we use a high-quality emotional dataset of 10,000 Mandarin speech samples from a single male speaker, where each utterance is human-annotated with one of five emotion categories (angry, happy, sad, surprised, and fearful). 
This same dataset is used for three purposes: 1) for the SFT, 2) for the corrective fine-tuning of the RM (Sec.~\ref{ssec:fine-tuning}), and 3) for the subsequent policy optimization (Sec.~\ref{ssec:rrpo}). This choice ensures objective consistency across all models and methods. The risk of overfitting is mitigated by the robust hybrid regularization scheme and the strong priors of the pre-trained model. 

Our RM architecture follows DiffRO~\cite{DiffRO}, with the primary improvement being a hybrid regularization scheme for the SER task head. 
The regularization hyperparameters are set to $\epsilon=0.1$, $\epsilon_\text{adv}=0.5$, and $\alpha=0.5$, while the learning rate is fixed at $1 \times 10^{-5}$ for all experiments. All models are trained on 8 NVIDIA A800 GPUs. 

\vspace{-0.2cm}
\subsubsection{Evaluation Metrics}
\label{sec:evaluation}

We evaluate our method using both subjective and objective metrics to demonstrate its effectiveness in mitigating reward hacking. 
The primary evidence for our claim comes from a subjective Mean Opinion Score (MOS) evaluation, as human perception serves as the ground truth for assessing whether a policy model has ``hacked" a reward model. 
For this assessment, we randomly sampled 10 utterances for each of the five emotion categories from a test set, which contains text prompts designed specifically for evaluation and not seen during training. 
These samples are evaluated by 20 native Mandarin speakers on two key metrics: 1) Emotion MOS (E-MOS), which assesses emotional expressiveness and adherence to the prompt, and 2) Naturalness MOS (N-MOS), which evaluates speech naturalness, including aspects like prosody and pronunciation. All ratings are scored on a 5-point scale from 1 to 5 with 0.5-point increments. 

To analyze the underlying reasons for this improved performance, we also conduct an objective ablation study on a downstream SER task. 
Our evaluation is based on the premise that an RM's robustness against reward hacking is directly related to its ability to generalize across diverse acoustic conditions and emotional styles. 
A model that performs well on varied, challenging data has likely learned a fundamental representation of emotion, making it less susceptible to exploitation. 
To test this, we evaluate our fine-tuned RM's generalization capabilities on several standard datasets~\cite{iemocap, ESD, mer2023}, which contain a mix of both acted and spontaneous emotional speech. 
Therefore, high performance across these varied domains serves as a strong objective indicator of the robustness required to prevent reward hacking.

\subsection{Experiments Results}
\label{ssec:result}

\begin{table}[!t]
    \centering
    \caption{Subjective evaluation results with 95\% confidence intervals. Higher is better. Best results are in \textbf{bold}.}
    \label{tab:mos}
    \begin{tabular}{l c c}
    \toprule
        \textbf{Method} & \textbf{E-MOS} ($\uparrow$) & \textbf{N-MOS} ($\uparrow$) \\
    \midrule
        CosyVoice2 (Baseline) & 3.27 ± 0.09 & 3.65 ± 0.06 \\
        + SFT & 3.52 ± 0.06 & 3.72 ± 0.07 \\
        \quad + DiffRO & 3.65 ± 0.11 & 3.61 ± 0.13 \\
        \quad + \textbf{RRPO (Ours)} & \textbf{3.78 ± 0.08} & \textbf{3.81 ± 0.09} \\
    \bottomrule
    \end{tabular}
    \vspace{-0.2cm}
\end{table} 

\begin{table}[!t]
    \centering
    \caption{Ablation study of our hybrid regularization (Sec.~\ref{ssec:fine-tuning}) on the reward model for the SER task. Results are Weighted Accuracy (\%). Higher is better.}
    \label{tab:ser_ablation}
    \begin{tabularx}{\columnwidth}{>{\RaggedRight}X c c c}
    \toprule
         \multirow{2}{*}{\textbf{Method}} 
         & \textbf{IEMOCAP} & \textbf{MER2023} & \textbf{ESD} \\ 
         & \textbf{(en)} & \textbf{(zh)} & \textbf{(zh)} \\
    \midrule
        DiffRO (Baseline) 
            & 66.0 & 50.9 & 64.4 \\
        + LS 
            & 66.8 & 51.4 & 72.8 \\
        \quad + EAM 
            & \textbf{69.1} & 52.7 & \textbf{82.3} \\
        \quad \quad + Adv (\textbf{RRPO}) 
            & 68.0 & \textbf{54.8} & 81.7 \\
    \bottomrule
    \end{tabularx}
    \vspace{-0.3cm}
\end{table}

\subsubsection{Subjective Evaluation Results}
\label{ssec:subjective}

Table~\ref{tab:mos} presents the results of our subjective evaluation. Our proposed RRPO method significantly outperforms all baselines, achieving the highest scores in both emotional expressiveness (E-MOS) and naturalness (N-MOS). 

Notably, the DiffRO baseline exhibits a behavior characteristic of reward hacking. While it achieves a competitive E-MOS score, its N-MOS score is substantially lower than even the SFT baseline. 
This discrepancy reveals that the policy model has found a detrimental ``shortcut" to maximize its reward. Instead of learning the complex task of generating genuine emotion, the policy model exploits vulnerabilities in the RM by producing subtle acoustic artifacts. These artifacts, while perceptually degrading to humans, provide a simple way to fool the RM into assigning a high score. 

In contrast, our RRPO method demonstrates a genuine and holistic improvement in perceptual quality, not only achieving the highest E-MOS but also surpassing the SFT baseline in N-MOS. 
This result confirms that our hybrid regularization scheme is effective in creating a robust RM whose reward signal is more reliably aligned with true human perception. 

\vspace{-0.2cm}
\subsubsection{Objective Evaluation and Ablation Study}
\label{ssec:subjective}

Table~\ref{tab:ser_ablation} presents our ablation study, showing that the full RRPO scheme significantly outperforms the DiffRO baseline across all datasets. 
The results reveal a nuanced interaction among the regularization components. 
EAM provides the most significant performance improvement from a single component, highlighting its strength in enhancing model generalization. 
However, while adding Adv further improves performance on MER2023, it slightly degrades performance on the other datasets. 
This is consistent with the known trade-off between generalization and adversarial robustness, suggesting Adv is most crucial for hardening the decision boundary against more challenging data distributions. 

The substantial improvement on the IEMOCAP dataset is particularly noteworthy, as it demonstrates strong cross-lingual generalization from Mandarin to English. 
Given that our RM was fine-tuned exclusively on Mandarin data, this result strongly suggests that our method compels the model to learn fundamental, language-agnostic representations of emotion, rather than relying on superficial, language-specific acoustic cues. 
This provides clear evidence of enhanced robustness, correcting the vulnerabilities of the baseline RM and yielding a more reliable reward signal that is significantly harder for a policy to ``hack".


\vspace{-0.2cm}
\subsubsection{Discussion}
\label{ssec:discussion}

The subjective and objective results together validate our central hypothesis. A standard RM often learns to reward spurious, non-human-like acoustic cues that are easy for a policy model to generate. 
Our robust RM, proven to generalize across diverse objective tasks, is less susceptible to these spurious correlations. This, in turn, compels the policy model to abandon simple shortcuts and instead learn the more complex acoustic features that align with genuine human perception. Therefore, by ensuring the RM's objective robustness, we effectively mitigate reward hacking and guide the policy model to generate outputs with superior subjective qualities. 

Furthermore, the generalizable nature of our hybrid regularization scheme suggests broader applicability. It can be adapted to improve the robustness of RMs for other acoustic attribute prediction tasks, such as audio quality or speaker identity. Moreover, the scheme can also be integrated into the large-scale pre-training phase, leveraging pseudo-labels to build a more robust foundational RM from the outset. This approach would improve the overall efficiency and scalability of reinforcement learning in the speech domain.

\section{Conclusion}
\label{sec:conclusion} 

In this paper, we propose RRPO, a RL framework for emotional TTS that addresses the critical challenge of reward hacking in differentiable reward optimization. 
The key innovation of this framework is a novel hybrid regularization scheme that produces a robust RM reliably aligned with human perception. 
Our experiments demonstrate that by using this robust RM, RRPO effectively mitigates reward hacking and achieves significant improvements in both the emotional expressiveness and naturalness of synthesized speech. 
For future work, the generalizable nature of our regularization scheme allows for its application to other  speech attributes and its integration into large-scale pre-training. 
This presents a scalable path toward creating more controllable and expressive TTS system, enhancing the broader applicability of RL in the speech domain.

\vfill\pagebreak
\bibliographystyle{IEEEbib}
\bibliography{refs}

\begin{thebibliography}{10}

\bibitem{cosyvoice2}
Zhihao Du, Yuxuan Wang, Qian Chen, Xian Shi, Xiang Lv, Tianyu Zhao, Zhifu Gao, Yexin Yang, Changfeng Gao, Hui Wang, et~al.,
\newblock ``Cosyvoice 2: Scalable streaming speech synthesis with large language models,''
\newblock {\em arXiv preprint arXiv:2412.10117}, 2024.

\bibitem{seedtts}
Philip Anastassiou, Jiawei Chen, Jitong Chen, Yuanzhe Chen, Zhuo Chen, Ziyi Chen, Jian Cong, Lelai Deng, Chuang Ding, Lu~Gao, et~al.,
\newblock ``Seed-tts: A family of high-quality versatile speech generation models,''
\newblock {\em arXiv preprint arXiv:2406.02430}, 2024.

\bibitem{emodiff}
Yiwei Guo, Chenpeng Du, Xie Chen, and Kai Yu,
\newblock ``Emodiff: Intensity controllable emotional text-to-speech with soft-label guidance,''
\newblock in {\em ICASSP 2023-2023 IEEE International Conference on Acoustics, Speech and Signal Processing (ICASSP)}. IEEE, 2023, pp. 1--5.

\bibitem{zetspeech}
Minki Kang, Wooseok Han, Sung~Ju Hwang, and Eunho Yang,
\newblock ``Zet-speech: Zero-shot adaptive emotion-controllable text-to-speech synthesis with diffusion and style-based models,''
\newblock in {\em Interspeech 2023}, 2023, pp. 4339--4343.

\bibitem{emomix}
Haobin Tang, Xulong Zhang, Jianzong Wang, Ning Cheng, and Jing Xiao,
\newblock ``Emomix: Emotion mixing via diffusion models for emotional speech synthesis,''
\newblock in {\em Interspeech 2023}, 2023, pp. 12--16.

\bibitem{voxinstruct}
Yixuan Zhou, Xiaoyu Qin, Zeyu Jin, Shuoyi Zhou, Shun Lei, Songtao Zhou, Zhiyong Wu, and Jia Jia,
\newblock ``Voxinstruct: Expressive human instruction-to-speech generation with unified multilingual codec language modelling,''
\newblock in {\em Proceedings of the 32nd ACM International Conference on Multimedia}, 2024, pp. 554--563.

\bibitem{emovoice}
Guanrou Yang, Chen Yang, Qian Chen, Ziyang Ma, Wenxi Chen, Wen Wang, Tianrui Wang, Yifan Yang, Zhikang Niu, Wenrui Liu, et~al.,
\newblock ``Emovoice: Llm-based emotional text-to-speech model with freestyle text prompting,''
\newblock {\em arXiv preprint arXiv:2504.12867}, 2025.

\bibitem{instructtts}
Dongchao Yang, Songxiang Liu, Rongjie Huang, Chao Weng, and Helen Meng,
\newblock ``Instructtts: Modelling expressive tts in discrete latent space with natural language style prompt,''
\newblock {\em IEEE/ACM Transactions on Audio, Speech, and Language Processing}, vol. 32, pp. 2913--2925, 2024.

\bibitem{preference}
Jinchuan Tian, Chunlei Zhang, Jiatong Shi, Hao Zhang, Jianwei Yu, Shinji Watanabe, and Dong Yu,
\newblock ``Preference alignment improves language model-based tts,''
\newblock in {\em ICASSP 2025-2025 IEEE International Conference on Acoustics, Speech and Signal Processing (ICASSP)}. IEEE, 2025, pp. 1--5.

\bibitem{emodpo}
Xiaoxue Gao, Chen Zhang, Yiming Chen, Huayun Zhang, and Nancy~F Chen,
\newblock ``Emo-dpo: Controllable emotional speech synthesis through direct preference optimization,''
\newblock in {\em ICASSP 2025-2025 IEEE International Conference on Acoustics, Speech and Signal Processing (ICASSP)}. IEEE, 2025, pp. 1--5.

\bibitem{DiffRO}
Changfeng Gao, Zhihao Du, and Shiliang Zhang,
\newblock ``Differentiable reward optimization for llm based tts system,''
\newblock in {\em Interspeech 2025}, 2025, pp. 2450--2454.

\bibitem{PPO}
John Schulman, Filip Wolski, Prafulla Dhariwal, Alec Radford, and Oleg Klimov,
\newblock ``Proximal policy optimization algorithms,''
\newblock {\em arXiv preprint arXiv:1707.06347}, 2017.

\bibitem{DPO}
Rafael Rafailov, Archit Sharma, Eric Mitchell, Christopher~D Manning, Stefano Ermon, and Chelsea Finn,
\newblock ``Direct preference optimization: Your language model is secretly a reward model,''
\newblock {\em Advances in neural information processing systems}, vol. 36, pp. 53728--53741, 2023.

\bibitem{GRPO}
Zhihong Shao, Peiyi Wang, Qihao Zhu, Runxin Xu, Junxiao Song, Xiao Bi, Haowei Zhang, Mingchuan Zhang, YK~Li, et~al.,
\newblock ``Deepseekmath: Pushing the limits of mathematical reasoning in open language models,''
\newblock {\em arXiv preprint arXiv:2402.03300}, 2024.

\bibitem{rewardhacking}
Dario Amodei, Chris Olah, Jacob Steinhardt, Paul Christiano, John Schulman, and Dan Man{\'e},
\newblock ``Concrete problems in ai safety,''
\newblock {\em arXiv preprint arXiv:1606.06565}, 2016.

\bibitem{gumbelsoftmax}
Eric Jang, Shixiang Gu, and Ben Poole,
\newblock ``Categorical reparameterization with gumbel-softmax,''
\newblock in {\em 2017 International Conference on Learning Representations (ICLR)}, 2017.

\bibitem{labelsmoothing}
Christian Szegedy, Vincent Vanhoucke, Sergey Ioffe, Jon Shlens, and Zbigniew Wojna,
\newblock ``Rethinking the inception architecture for computer vision,''
\newblock in {\em 2016 IEEE Conference on Computer Vision and Pattern Recognition (CVPR)}. IEEE, 2016, pp. 2818--2826.

\bibitem{Wavlm}
Sanyuan Chen, Chengyi Wang, Zhengyang Chen, Yu~Wu, Shujie Liu, Zhuo Chen, Jinyu Li, Naoyuki Kanda, Takuya Yoshioka, Xiong Xiao, et~al.,
\newblock ``Wavlm: Large-scale self-supervised pre-training for full stack speech processing,''
\newblock {\em IEEE Journal of Selected Topics in Signal Processing}, vol. 16, no. 6, pp. 1505--1518, 2022.

\bibitem{LQF}
Qifei Li, Yingming Gao, Yuhua Wen, Ziping Zhao, Ya~Li, and Björn~W. Schuller,
\newblock ``Seenet: A soft emotion expert and data augmentation method to enhance speech emotion recognition,''
\newblock {\em IEEE Transactions on Affective Computing}, vol. 16, no. 3, pp. 2142--2156, 2025.

\bibitem{mixup}
Hongyi Zhang, Moustapha Cisse, Yann~N. Dauphin, and David Lopez-Paz,
\newblock ``mixup: Beyond empirical risk minimization,''
\newblock in {\em 2018 International Conference on Learning Representations (ICLR)}, 2018.

\bibitem{FGM}
Takeru Miyato, Andrew~M Dai, and Ian Goodfellow,
\newblock ``Adversarial training methods for semi-supervised text classification,''
\newblock in {\em 2017 International Conference on Learning Representations (ICLR)}, 2017.

\bibitem{iemocap}
Carlos Busso, Murtaza Bulut, Chi-Chun Lee, Abe Kazemzadeh, Emily Mower, Samuel Kim, Jeannette~N Chang, Sungbok Lee, and Shrikanth~S Narayanan,
\newblock ``Iemocap: Interactive emotional dyadic motion capture database,''
\newblock {\em Language resources and evaluation}, vol. 42, no. 4, pp. 335--359, 2008.

\bibitem{ESD}
Kun Zhou, Berrak Sisman, Rui Liu, and Haizhou Li,
\newblock ``Emotional voice conversion: Theory, databases and esd,''
\newblock {\em Speech Communication}, vol. 137, pp. 1--18, 2022.

\bibitem{mer2023}
Zheng Lian, Haiyang Sun, Licai Sun, Kang Chen, Mingyu Xu, Kexin Wang, Ke~Xu, Yu~He, Ying Li, Jinming Zhao, et~al.,
\newblock ``Mer 2023: Multi-label learning, modality robustness, and semi-supervised learning,''
\newblock in {\em Proceedings of the 31st ACM international conference on multimedia}, 2023, pp. 9610--9614.

\end{thebibliography}
\end{document}